\documentclass[useAMS,usegraphicx,usenatbib]{mn2e}

\newcommand{\de}{$\degr$}
\newcommand{\kms}{\hbox{km s$^{-1}$}}
\newcommand{\vsini}{\hbox{$v\,\sin\,i$}}

\title[Doppler images of SZ Psc]{The first Doppler images of the eclipsing binary \hbox{SZ Piscium}}
\author[Y.~Xiang, S.~Gu, A.~Collier~Cameron, J.~R.~Barnes and L.~Zhang]{Yue Xiang,$^{1}$$^{2}$$^{3}$\thanks{E-mails: xy@ynao.ac.cn (YX); shenghonggu@ynao.ac.cn(SG)} Shenghong Gu,$^{1}$$^{2}$$^{\star}$ A.~Collier~Cameron,$^{4}$ J.~R.~Barnes$^{5}$ \and and Liyun Zhang$^{6}$\\
$^{1}$Yunnan Observatories, Chinese Academy of Sciences, Kunming 650011, China\\
$^{2}$Key Laboratory for the Structure and Evolution of Celestial Objects, Chinese Academy of Sciences, Kunming 650011, China\\
$^{3}$University of Chinese Academy of Sciences, Beijing 100049, China\\
$^{4}$School of Physics and Astronomy, University of St Andrews, Fife KY16 9SS, UK\\
$^{5}$Department of Physical Sciences, The Open University, Walton Hall, Milton Keynes MK7 6AA, UK\\
$^{6}$Department of Physics, College of Science, Guizhou University and NAOC-GZU-Sponsored Center for Astronomy, Guizhou University,\\ Guiyang 550025, China}

\begin{document}

\maketitle

\label{firstpage}

\begin{abstract}
We present the first Doppler images of the active eclipsing binary system SZ Psc, based on the high-resolution spectral data sets obtained in 2004 November and 2006 September--December. The least-squares deconvolution technique was applied to derive high signal-to-noise profiles from the observed spectra of SZ Psc. Absorption features contributed by a third component of the system were detected in the LSD profiles at all observed phases. We estimated the mass and period of the third component to be about $0.9 M_{\odot}$ and $1283 \pm 10$ d, respectively. After removing the contribution of the third body from the LSD profiles, we derived the surface maps of SZ Psc. The resulting Doppler images indicate significant starspot activities on the surface of the K subgiant component. The distributions of starspots are more complex than that revealed by previous photometric studies. The cooler K component exhibited pronounced high-latitude spots as well as numerous low- and intermediate-latitude spot groups during the entire observing seasons, but did not show any large, stable polar cap, different from many other active RS CVn-type binaries.
\end{abstract}


\section{Introduction}

SZ Psc is a double-lined partial eclipsing binary composed of an F8V hotter and a K1IV cooler component, with an orbital period of about 3.97 d. The cooler component is larger and more massive than the hotter one and has filled 85\% of its Roche lobe \citep{pop1988}. The rotation of the hotter component is several times slower than its synchronous value, while the cooler component shows synchronous rotation \citep{eaton2007,gla2008}. SZ Psc is very active and classified as a member of RS CVn-type stars \citep{hall1976}. It shows strong chromospheric emission lines attributed to its cooler component \citep{ram1981,pop1988,doyle1994,fra1994}. Starspot activities on the K star were also revealed by many photometric studies \citep{eaton1979,lan2001,kang2003,eaton2007}. The orbital period of SZ Psc is not constant \citep{jak1976,tunca1984,kal1995}, which is similar to those of many other active binary systems. \citet{kal1995} derived a periodicity of 56 yr and an amplitude of $4.3\times10^{-4}$ d for the period change of the system. They suggested that it can be explained by a combination of the magnetic activity and the stellar wind.

SZ Psc is suspected to be a triple system. \citet{eaton2007} revealed that the systemic velocity of the binary is changing with time, which indicates a third component in SZ Psc. They suggested an amplitude less than 8 \kms\ and a period of 1143 or 1530 d for the systemic velocity and thus inferred that the third component is a cool dwarf with a mass of about 0.9--1.0$M_{\odot}$. They also found weak features in the \mbox{Na~{\sc i}} D lines probably contributed by the third component and estimated its contribution to be about 3\%--4\% of the brightness of SZ Psc. So far, the physical properties of the third component and the outer orbit of SZ Psc are still poorly known.

\citet{zhang2008} analysed several chromospheric activity indicators using the spectral subtraction technique and revealed the rotational modulation of the activity on the cooler component of SZ Psc. In addition, they found absorption features in the H$_{\alpha}$ profiles of SZ Psc probably accounted for by prominence-like material around the K star or mass transfer between two components. Using higher time-resolved spectra, \citet{cao2012} also detected absorption features in the H$_{\alpha}$ profiles, which indicate prominence activity on the cooler component. Their calculation shows that the distance of the prominence from the K star's rotation axis exceeded the Roche lobe of the K star.

\citet{lan2001} derived surface images of both components of SZ Psc from long-term photometric observations. They revealed the presence of several active regions on the surface of the cooler component of SZ Psc. One of them is stable and facing the hotter component. \citet{kang2003} derived unique solutions from light curves with good phase sampling using the starspot model and revealed that the variations of the shape of light curves are mainly accounted for by spot evolution and migration on the K star of SZ Psc. \citet{eaton2007} suggested that the cooler component have many small starspots rather than a few large ones, because its line profiles lack large distortions.

In order to investigate the starspot activities on active close binaries, we have carried out a series of high-resolution spectroscopic observations on targets with various stellar parameters and evolutionary stages \citep{gu2003,xiang2014,xiang2015}. In this work, we have derived the surface images of the K subgiant component of SZ Psc for 2004 November, 2006 September, October, November and December, through Doppler imaging technique. To our knowledge, there is no Doppler image for SZ Psc before, which could offer us a more detailed distribution of starspots than light-curve modelling. We shall describe the observations and data reduction in Section 2. The Doppler images will be given and discussed in Section 3 and 4, respectively. In Section 5, we shall summarize the present work.

\section{Observations and data reduction}

\begin{table}
 \caption{The full version of the observing log.}
 \label{tab:log}
 \begin{tabular}{lcccc}
  \hline
  UT Date   & HJD & Exp. & S/N & S/N\\
  &2450000+&(s)&Input&LSD\\
  \hline
  20/11/2004 & 3330.1055 & 2400 & 141 & 1911\\
  20/11/2004 & 3330.1365 & 2400 & 134 & 1818\\
  21/11/2004 & 3331.1073 & 2400 &  87 & 1180\\
  27/11/2004 & 3337.1254 & 2400 & 103 & 1407\\
  01/09/2006 & 3980.1557 & 1800 &  55 &  750\\
  01/09/2006 & 3980.1769 & 1800 &  62 &  847\\
  04/09/2006 & 3983.1880 & 2100 &  83 & 1131\\
  04/09/2006 & 3983.2125 & 2100 &  92 & 1262\\
  05/09/2006 & 3984.1295 & 1800 & 101 & 1377\\
  05/09/2006 & 3984.1505 & 1800 & 105 & 1430\\
  06/09/2006 & 3985.1228 & 1800 & 121 & 1654\\
  06/09/2006 & 3985.1437 & 1800 & 130 & 1776\\
  28/10/2006 & 4036.9567 & 1800 & 131 & 1794\\
  28/10/2006 & 4036.9777 & 1800 & 132 & 1811\\
  28/10/2006 & 4037.0227 & 1800 & 142 & 1948\\
  28/10/2006 & 4037.0438 & 1800 & 149 & 2039\\
  28/10/2006 & 4037.0836 & 1800 & 135 & 1849\\
  28/10/2006 & 4037.1051 & 1800 & 127 & 1741\\
  28/10/2006 & 4037.1758 & 1800 & 106 & 1448\\
  28/10/2006 & 4037.1968 & 1800 & 112 & 1530\\
  29/10/2006 & 4037.9357 & 1800 &  85 & 1162\\
  29/10/2006 & 4037.9568 & 1800 &  88 & 1199\\
  29/10/2006 & 4037.9901 & 1800 &  85 & 1156\\
  29/10/2006 & 4038.0125 & 1800 &  83 & 1136\\
  29/10/2006 & 4038.0335 & 1800 &  87 & 1188\\
  29/10/2006 & 4038.0551 & 1800 &  89 & 1216\\
  30/10/2006 & 4039.0799 & 1800 &  95 & 1295\\
  28/11/2006 & 4067.9843 & 1500 &  84 & 1152\\
  28/11/2006 & 4068.0029 & 1500 &  90 & 1234\\
  28/11/2006 & 4068.1349 & 2400 &  95 & 1298\\
  29/11/2006 & 4068.9424 & 1500 &  84 & 1141\\
  29/11/2006 & 4068.9599 & 1500 &  83 & 1135\\
  29/11/2006 & 4069.0729 & 1500 &  86 & 1167\\
  29/11/2006 & 4069.0905 & 1500 &  81 & 1105\\
  29/11/2006 & 4069.1080 & 1500 &  72 &  984\\
  29/11/2006 & 4069.1311 & 2100 &  77 & 1049\\
  30/11/2006 & 4069.9889 & 3000 & 126 & 1713\\
  30/11/2006 & 4070.1327 & 3000 & 118 & 1606\\
  01/12/2006 & 4070.9555 & 3600 &  58 &  794\\
  01/12/2006 & 4071.0433 & 3600 &  48 &  669\\
  07/12/2006 & 4077.0996 & 3600 &  85 & 1164\\
  08/12/2006 & 4077.9564 & 3600 &  96 & 1314\\
  08/12/2006 & 4078.1182 & 3600 &  45 &  621\\
  09/12/2006 & 4078.9482 & 3600 & 105 & 1434\\
  09/12/2006 & 4079.1085 & 3600 & 109 & 1488\\
  10/12/2006 & 4079.9563 & 3600 & 146 & 1991\\
  10/12/2006 & 4080.1041 & 3600 & 118 & 1615\\
  11/12/2006 & 4080.9480 & 3600 & 121 & 1652\\
  11/12/2006 & 4081.0727 & 3000 &  86 & 1170\\
  11/12/2006 & 4081.1132 & 2400 &  55 &  751\\
  \hline
 \end{tabular}
\end{table}

The observations of SZ Psc were carried out on 2004 November 20--27, 2006 September 01--06, 2006 October 28--30, 2006 November 28--December 01 and 2006 December 07--11, using the Coud\'{e} echelle spectrograph \citep{zhao2001} mounted on the 2.16m telescope at the Xinglong station of the National Astronomical Observatories of China. In all observations, a 1024$\times$1024 pixel TEK CCD detector was deployed to record the data. As results, the spectral region was about 5600--9000 \AA\ and the resolution was R = 37000. The observations are the same as that of \citet{zhang2008}, but the number of spectra we used is more. Their work aimed to investigate chromospheric activities of the cooler component by analysing several indicators, so the data obtained at the eclipsing phases were omitted and some spectra were combined to improve the signal-to-noise ratios. In our case, since the imaging code can handle the eclipse and we can obtain enough S/N for a single observation through the LSD calculation, we used all of available spectral data to derive Doppler images for SZ Psc.

The observing log, including the UT date, Heliocentric Julian date, exposure time, is given in Table \ref{tab:log}, which is available online. In the next section, we will use the spectral data obtained on 2006 October 28, 29 and 30 to show signatures of the third component, but the last two spectra observed on 2006 October 28 will be excluded because the lines of the F star and the third component nearly coincided.

Apart from SZ Psc, we also observed three slowly-rotating, inactive template stars, HR 6669 (F8V), HR 7948 (K1IV) and HR 248 (M0III), by using the same instrument setup as SZ Psc, to mimic the local intensity profiles of the photosphere and starspot for each component of SZ Psc, which are required for the two-temperature model of our imaging code.

The spectroscopic data were reduced using the IRAF\footnote{IRAF is distributed by the National Optical Astronomy Observatory, which is operated by the Association of Universities for Research in Astronomy (AURA) under cooperative agreement with the National Science Foundation.} package in a standard way. The reduction procedure included image trimming, bias subtraction, flat-field dividing, scatter light subtraction, cosmic-ray removal, 1D spectrum extraction, wavelength calibration and continuum fitting. The wavelength calibration was carried out using the comparison spectra of the ThAr lamp taken at each night.

Since starspot signatures are very small, we enhanced the S/N of the observed profiles by applying the least-squares deconvolution (LSD; \citealt{don1997}) technique, which combines all available atomic lines in one observed spectrum into a single mean profile. We derived the line list, which contained central wavelength and line depth, for the model atmosphere with $T_{eff}$ = 5000 K, from the Vienna Atomic Line Database (VALD; \citealt{kup1999}). The wavelength regions of strong telluric and chromospheric lines were excluded from the line list to prevent their influence. Then we calculated the LSD profile from each observed spectrum with the line list. In the LSD calculations we set the increment per pixel to 4.1 \kms, according to the resolution power of our spectral data.  We list the peak S/N of the observed spectra and the S/N of the corresponding LSD profiles in Table \ref{tab:log}. The S/N gain is around 14 for our observations. To correct the errors in radial velocity caused by the instrumental shift, we take advantage of the telluric lines in the spectra, which have negligible velocities, as described by \citet{cam1999}. A LSD profile was derived using the list of telluric lines for each spectrum; the first one in each observing run was used as the template. Then the shift derived from the cross-correlation between the telluric LSD profile and the template was corrected. As demonstrated by \citet{don2003}, this method can provide a precision better than 0.1 \kms.

The spectra of the template stars of the photospheres and starspots were also deconvolved in the same manner to produce the lookup tables. The linear limb-darkening coefficients for UBVRI passbands derived by \citet{cla2011} were used to obtain the values at the centroidal wavelength of the LSD profile for the photospheres and starspots. Then 30 limb angles were used for producing the lookup tables.

\begin{figure}
\centering
\includegraphics[width=0.45\textwidth]{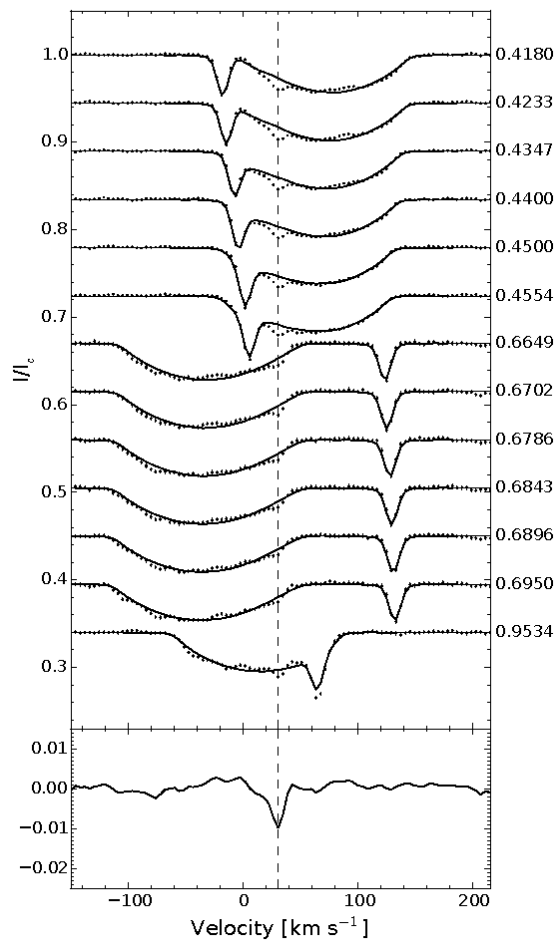}
\caption{The upper panel shows observed LSD profiles (dots) and calculated profiles of immaculate photosphere (solid lines). The spectra at the orbital phases when the lines of the hotter component and the third one were not blurred significantly were selected. The lower panel shows the mean residual between observed and calculated profiles. The vertical dashed line shows the positions of the absorption features, which are probably contributed by the third component of SZ Psc.}
\label{fig:third_body}
\end{figure}

\begin{figure}
\centering
\includegraphics[angle=270,width=0.475\textwidth]{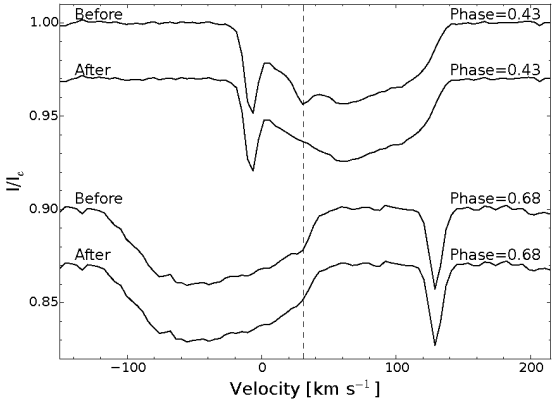}
\caption{Examples for removing the contribution of the third component. The dashed line shows the position of the profile of the third component.}
\label{fig:example}
\end{figure}

\section{Doppler imaging}

\subsection{The third component}

The LSD profiles of SZ Psc with high S/N offer us an opportunity to detect the spectral line of the third body directly. For example, we select the LSD profiles of the 2006 October data set at the orbital phases when the absorption lines of the F star and the possible third one are not blurred seriously, and plot them as well as the corresponding calculated profiles that represent the immaculate photospheres of the binary in Fig. \ref{fig:third_body}. As shown in this figure, the possible line of the third star was blurred by the highly rotational broadened profile of the cooler component, which is deformed by the starspots significantly, at all observed phases. According to \citet{eaton2007}, the period of the triple system is very long, thus we may expect that the velocity of the third body changed little in one observing run relative to the resolution of our spectral data (4.1 \kms). On the other hand, the deformation of profiles caused by starspots is changing with orbital phase. Therefore we derive the mean of the residuals between the observed and the calculated profiles to reduce the effect of starspot distortions. The result is shown in the bottom panel of Fig. \ref{fig:third_body}. As seen in the figure, clear absorption features were detected in the LSD profiles of SZ Psc, which confirmed that SZ Psc is a triple system.

Since the weak features may produce the same effect as starspot distortions, we subtracted the contribution of the third component assuming a Gaussian profile. Firstly, we estimated the radial velocity of the third body by cross-correlating the residuals with the template telluric profile for each observing run. In our case, we assumed that the intrinsic velocity of the third body was fixed in one observing run and only changed the velocity depending on the heliocentric motion for each observed phase. Then we created a Gaussian profile and multiply it by a factor depending on the brightness of the binary system at different phases. We also used imaging code to fine tune the profile by minimizing $\chi^{2}$ as described in the next subsection. Examples for removing the contribution of the third component are shown in Fig. \ref{fig:example}. The radial velocity of the third star determined for each observing run is given in the second column of Table \ref{tab:velocity}.

\begin{table}
 \caption{The radial velocities of the third star and the binary system, derived from five observing runs. The errors are 0.4 and 0.6 \kms, respectively.}
 \label{tab:velocity}
 \begin{tabular}{lcc}
  \hline
  Date & RV$_{third}$ & $\gamma$\\
       &  \kms       &  \kms\\
  \hline
  2004 Nov.  & 19.0 & 8.5\\
  2006 Sept. & 3.1  & 12.7\\
  2006 Oct.  & 8.4  & 11.3\\
  2006 Nov.  & 11.4 & 10.2\\
  2006 Dec.  & 12.9 & 9.7\\
  \hline
 \end{tabular}
\end{table}

\subsection{System parameters}

Accurate stellar parameters are required to derive reliable Doppler images and prevent from reconstructing artefacts \citep{cam1994}, especially for eclipsing binary systems \citep{vin1993}. It has been demonstrated that the imaging code can be used to determine stellar parameters for both of single and binary stars \citep{bar1998,bar2004}. Fine tuning stellar parameters can be achieved by performing a fixed number of maximum entropy iterations with various combinations of parameters and then finding the best-fit values that leads to a minimum $\chi^{2}$. The advantage of this method is that the effect of starspot distortions on parameter determinations is removed \citep{bar2005}.

In our case, firstly, we took the improved parameters, such as the radial velocity curve amplitudes of the F hotter ($K_{1}$) and the K cooler ($K_{2}$) components, the inclination of the orbital axis ($i$), the conjunction time ($T_{0}$; the hotter component is behind) and the orbital period ($P$), from the paper by \citet{eaton2007}, who used both of the photometric and spectroscopic data; we did not change these values in the following steps. Then we used $\chi^{2}$ minimization method to determine the projected rotational velocity (\vsini) of each component, from the combined data set of 2006 November and December; and the systematic radial velocity of the binary ($\gamma$), from each data set. We list the final adopted values for imaging SZ Psc in Table \ref{tab:par}, except for the values of $\gamma$ for five observing runs, which are given in the third column of Table \ref{tab:velocity} for comparison. The rotational velocity of the cooler component we derived is obviously lower than that (80 \kms) determined by \citet{eaton2007} but very close to the value (70 \kms) derived by \citet{str1993} and \citet{gla2008}.

\begin{table}
 \caption{Adopted stellar parameters of SZ Psc for Doppler imaging. The F hotter component is defined as the primary and the K cooler star is the secondary.}
 \label{tab:par}
 \begin{tabular}{lcc}
 \hline
 Parameter & Value & Ref.\\
 \hline
 $q=M_{2}/M_{1}$ & 1.40 & a\\
 $K_{1}$ (km s$^{-1}$) & 103.98 & a\\
 $K_{2}$ (km s$^{-1}$) & 74.2 & a\\
 $i$ (\de) & 69.75 & a\\
 $T_{0}$ (HJD) & 2449284.4483 & a\\
 $P$ (d)  & 3.96566356 & a\\
 \vsini~$_{1}$ (km s$^{-1}$) & $3.0 \pm 0.6$ & DoTS\\
 \vsini~$_{2}$ (km s$^{-1}$) & $67.7 \pm 1.0$ & DoTS\\
 $T_{eff,1}$ (K)& 6090 & b\\
 $T_{eff,2}$ (K)& 4910 & b\\
 albedo$_{1}$ & 1.0 & b\\
 albedo$_{2}$ & 0.3 & b \\
 \hline
 \end{tabular}\\
  References: a. \citet{eaton2007}; b. \citet{lan2001}.\\
\end{table}

\subsection{Results}

\begin{figure*}
\centering
\includegraphics[angle=270,width=0.95\textwidth]{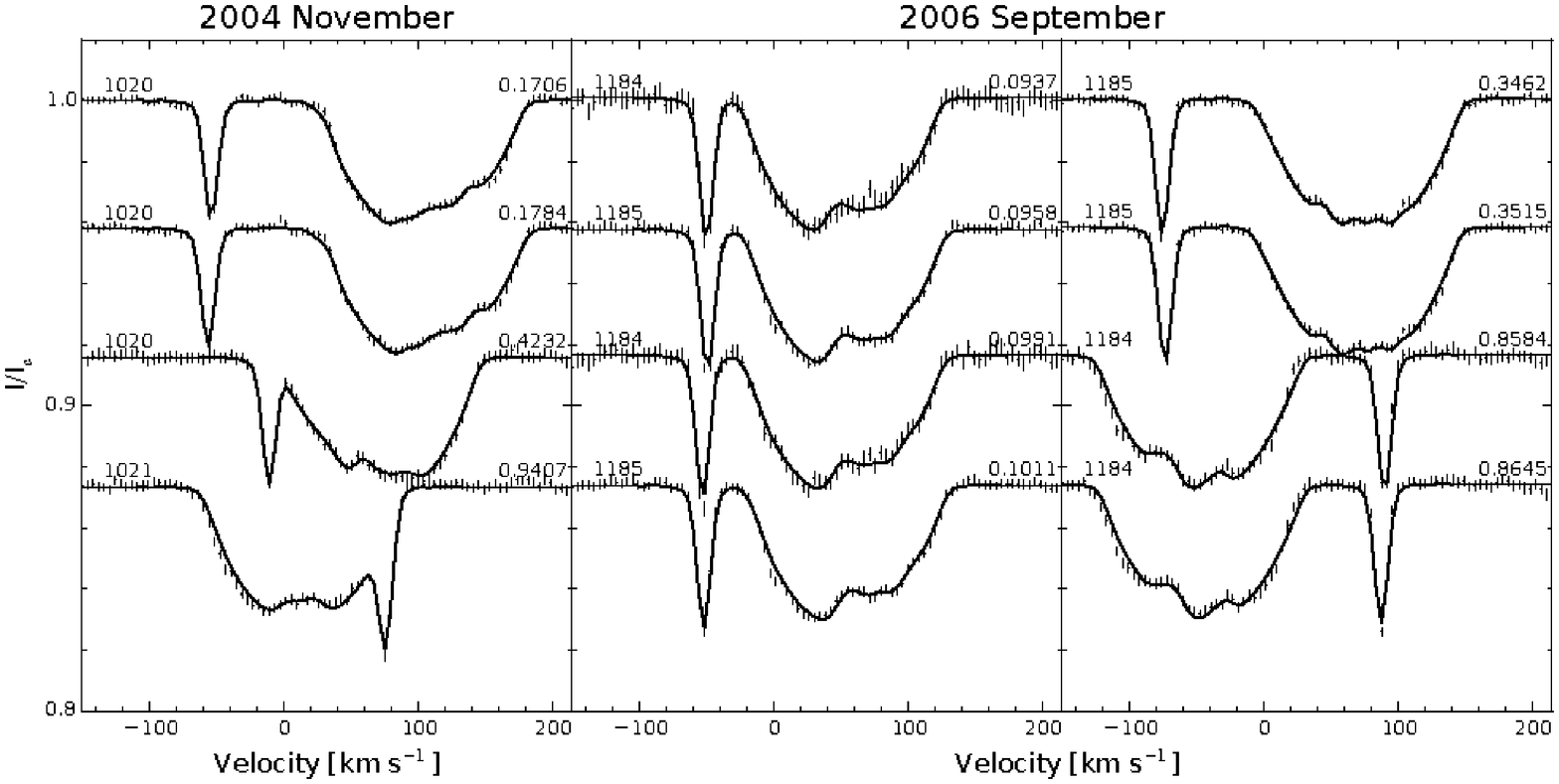}
\includegraphics[angle=270,width=0.95\textwidth]{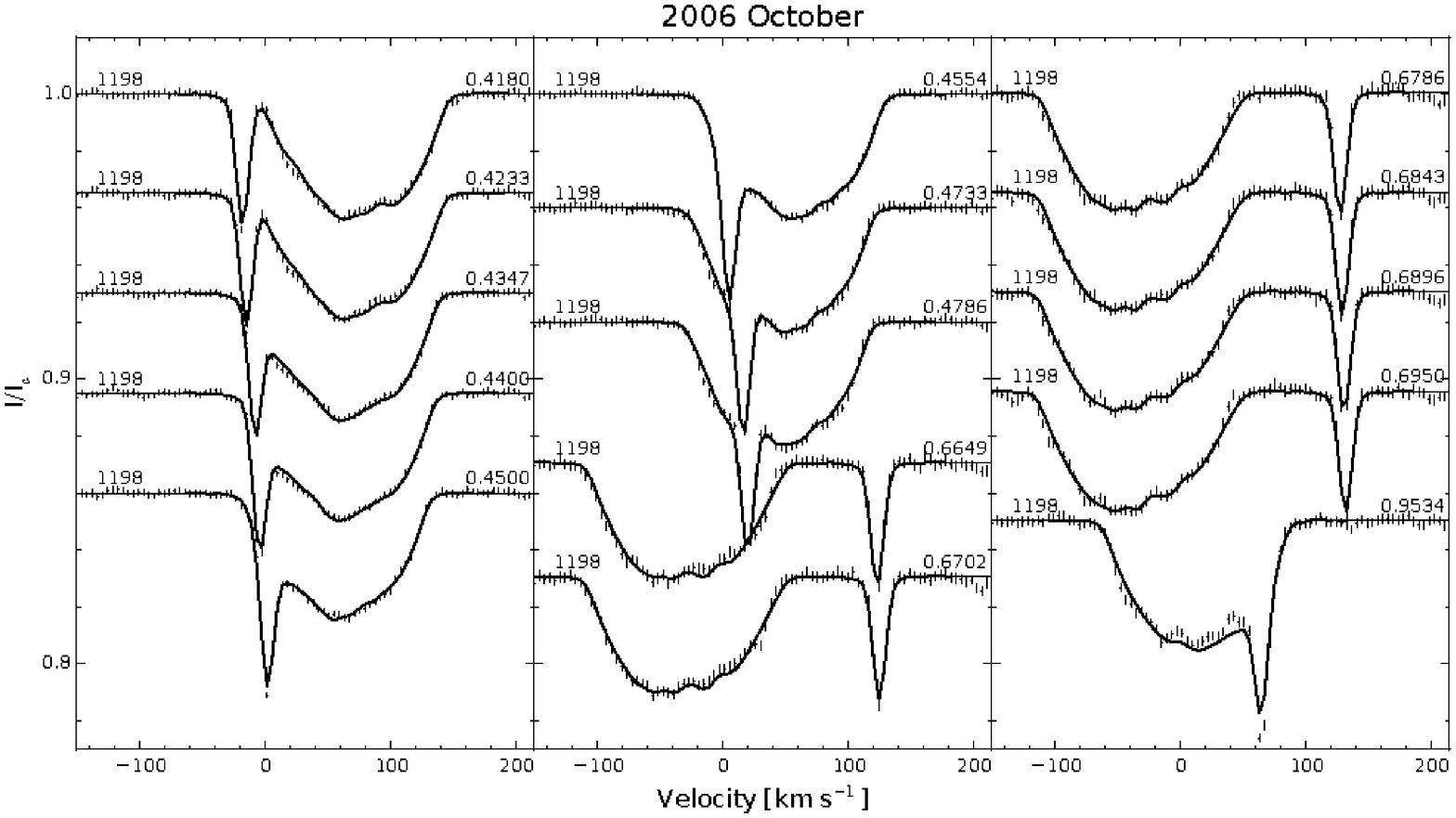}
\caption{The observed LSD profiles (dots with error bars) and the maximum entropy solutions (solid lines) for 2004 November, 2006 September and 2006 October. The rotation number and orbital phase are marked at the left and right side of each profile, respectively.}
\label{fig:spec1}
\end{figure*}

\begin{figure*}
\centering
\includegraphics[angle=270,width=0.9\textwidth]{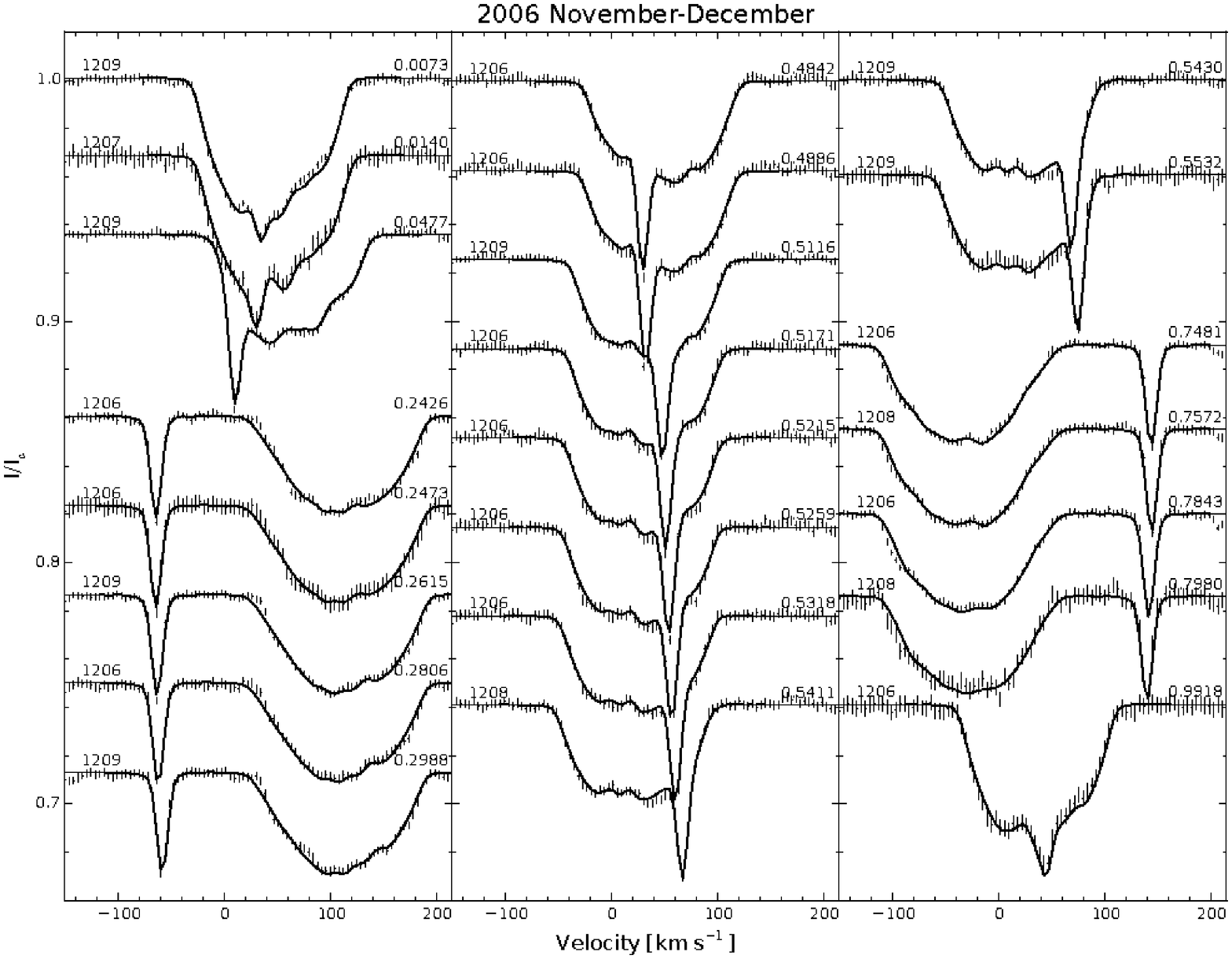}
\caption{Same as Fig. \ref{fig:spec1}, but for 2006 November--December.}
\label{fig:spec2}
\end{figure*}

\begin{figure*}
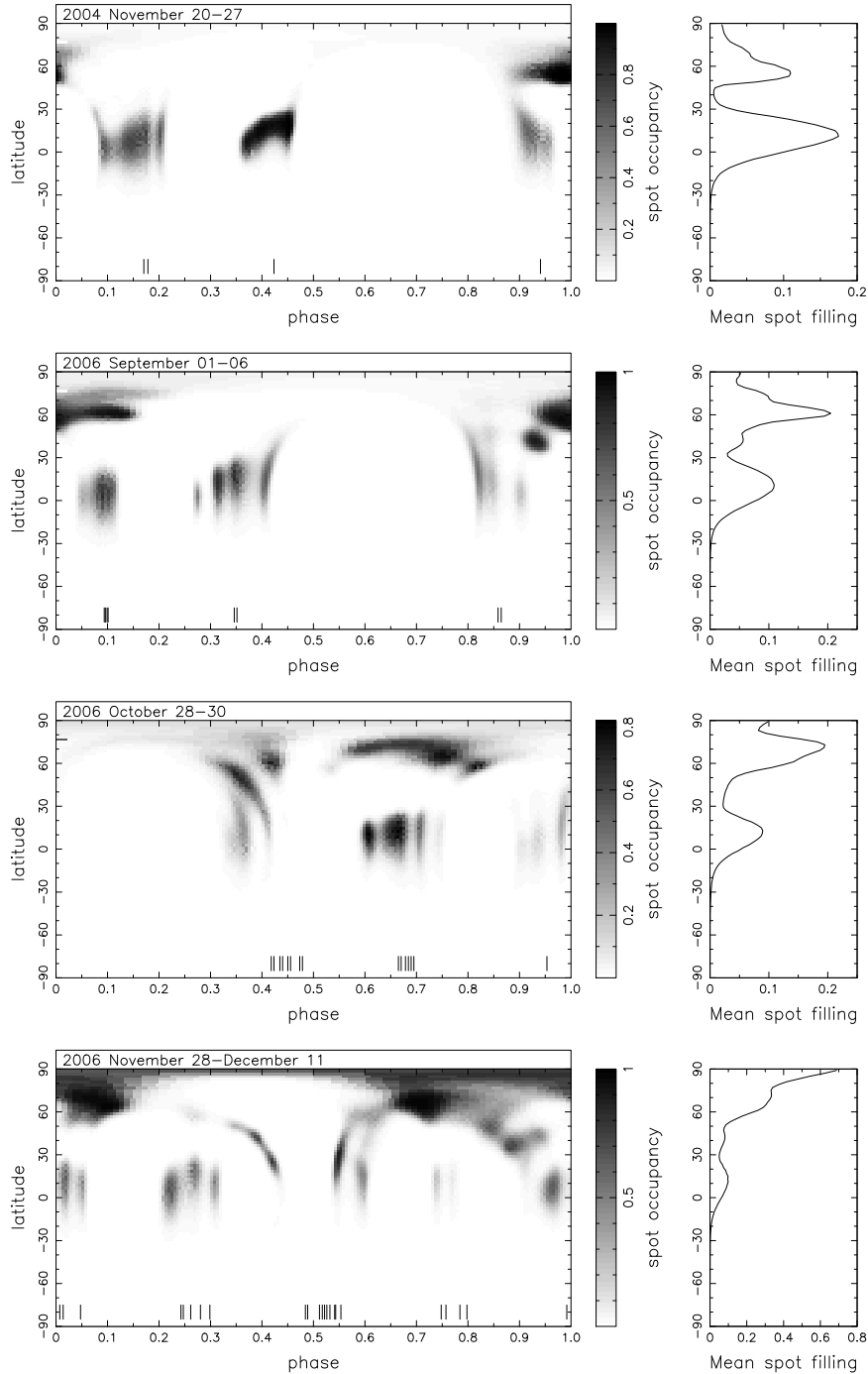

\centering
\includegraphics[angle=270,width=0.66\textwidth]{i0411.eps}
\includegraphics[angle=270,width=0.66\textwidth]{i0609.eps}
\includegraphics[angle=270,width=0.66\textwidth]{i0610.eps}
\includegraphics[angle=270,width=0.66\textwidth]{i061112.eps}
\caption{Mercator projection of reconstructed images of 2004 November, 2006 September, 2006 October, and 2006 November--December (from top to bottom). Observing phases are marked by ticks. Mean spot filling factor as a function of latitude is also plotted at the right side of each surface image.}
\label{fig:mercator}
\end{figure*}

We used the imaging code Doppler Tomography of Star (DoTS; \citealt{cam1992}; \citealt{cam1997}) to perform the maximum entropy regularized iterations to each data set. Minor modifications were made on the imaging code to take into account the non-synchronous rotation of the hotter component of SZ Psc. The vertexes of the surface grid of the hotter component were shifted in each observed phase and given new radial velocities when projected onto the viewplane, depending on the rotational velocity of the F star. The observed LSD profiles and the corresponding maximum entropy solutions are plotted in Fig. \ref{fig:spec1} and \ref{fig:spec2}; the Mercator projection of each resulting surface image for the cooler component of SZ Psc is presented in Fig. \ref{fig:mercator}. Note that phase 0.5 on the cooler component faces the hotter component in our images.

In 2004 November, the cooler component of SZ Psc showed significant low-latitude spot patterns around phases 0.15, 0.4 and a pronounced high-latitude spot around phase 0. The image of 2006 September shows an extended intermediate-latitude spot in phase 0.9--1.1. Meanwhile, low latitudes also show the presence of several active regions around phases 0.1 and 0.35. In 2006 October, a pronounced low-latitude spot group appeared between phases 0.6 and 0.7. The image of 2006 November--December indicates pronounced high-latitude spot patterns located at phases 0.1 and 0.75. Besides, there are several low- and intermediate-latitude spot groups around phases 0, 0.25, 0.4, 0.55 and 0.9. In order to show the relationship between the starspot distribution and the relative positions of two components of SZ Psc clearly, we present the images of the binary system at phases 0, 0.25, 0.5 and 0.75, of one orbital cycle in Fig. \ref{fig:images}, using the Doppler image of 2006 November--December. In this figure, we can also see the non-synchronous rotation of the F star from its starspot rotation. Since the rotational velocity of the hotter component is very small, the spots on it are hardly resolved and mainly produce the changes of the line strength of the F star at different orbital phases.

Poor phase sampling may lead to artefacts in the reconstructed images. Hence we tested for the reliability of our Doppler images, as discussed in Appendix A (online only). The results indicate that arc-shaped features in all of the surface images of the K star are spurious, especially for large phase gaps, where show pronounced arc-shaped artefacts (e.g. the prominent features at phases 0.4 and 0.8 in 2006 September). In addition, starspots at unobserved phases may be absent in the reconstructed images.

\section{Discussion}

\begin{figure}
\centering
\includegraphics[angle=270,width=0.35\textwidth]{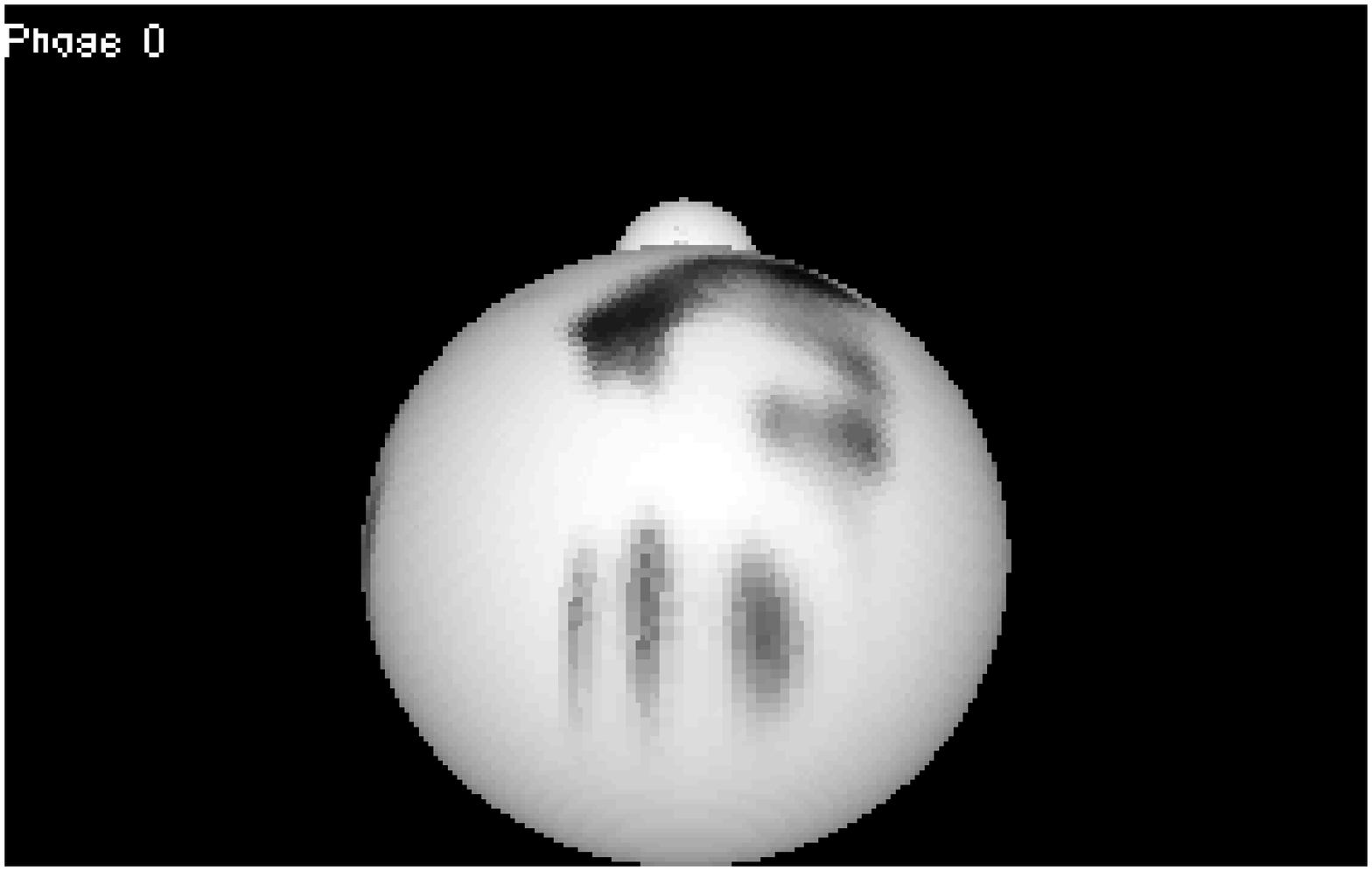}
\includegraphics[angle=270,width=0.35\textwidth]{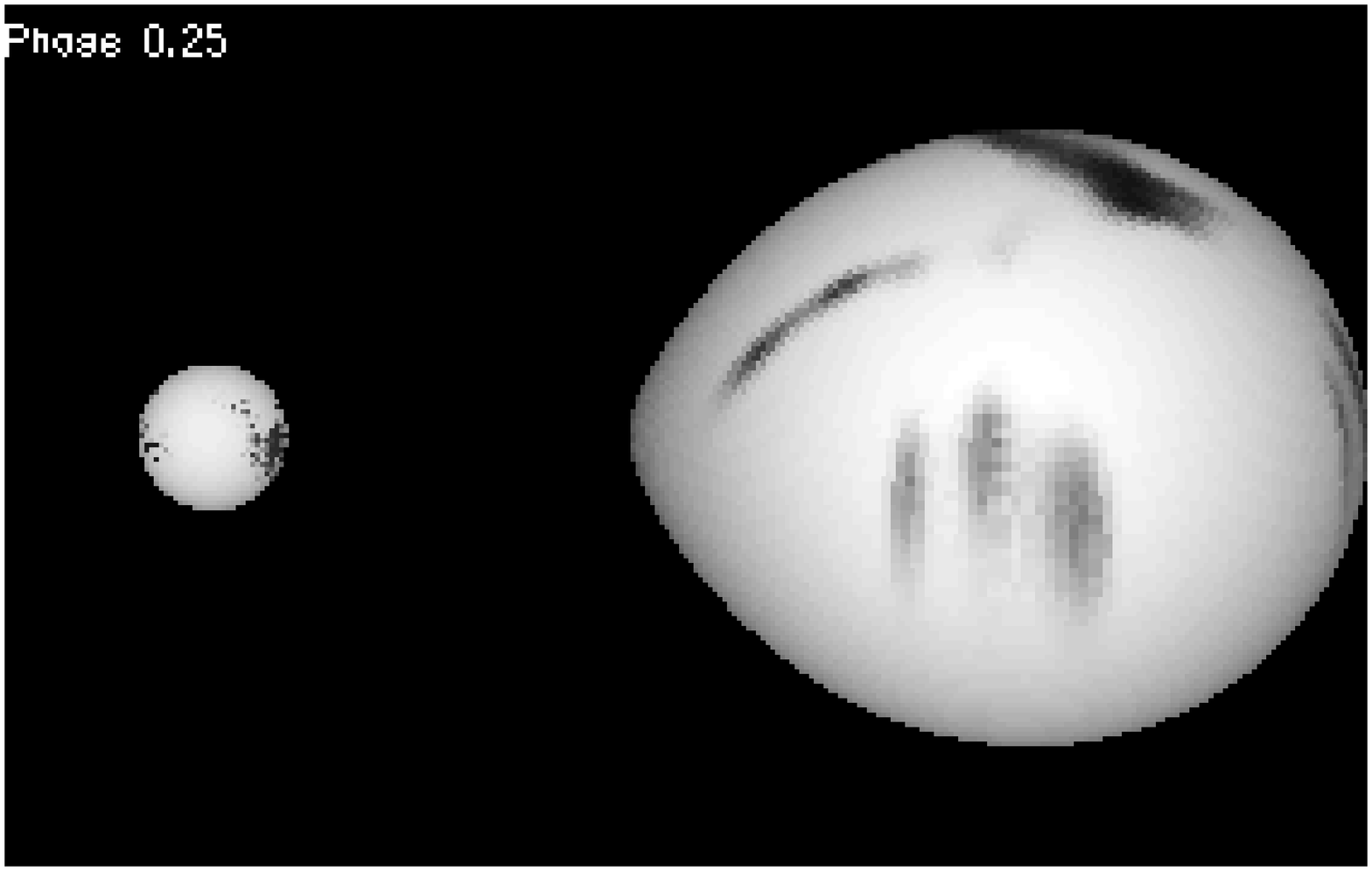}
\includegraphics[angle=270,width=0.35\textwidth]{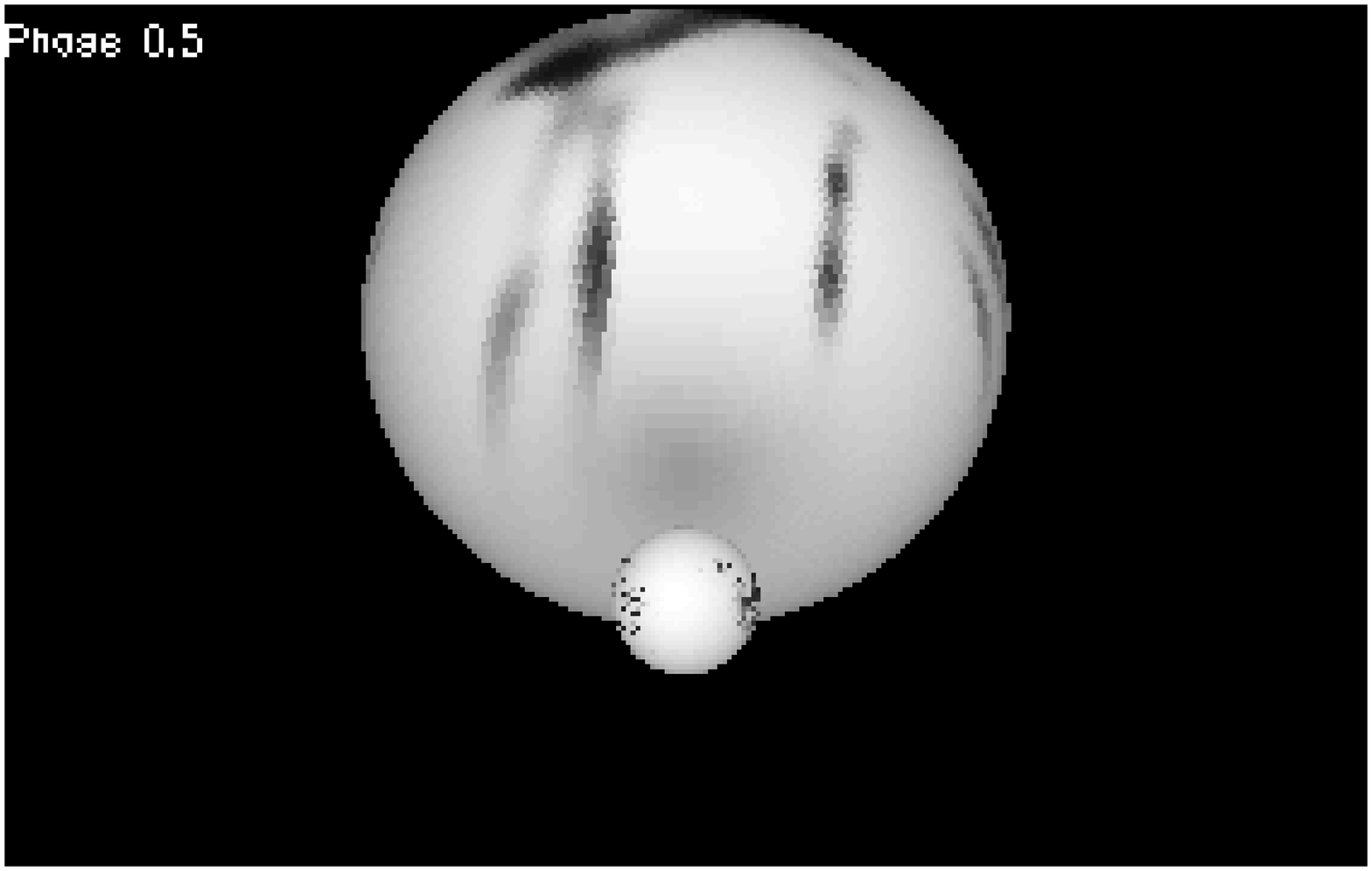}
\includegraphics[angle=270,width=0.35\textwidth]{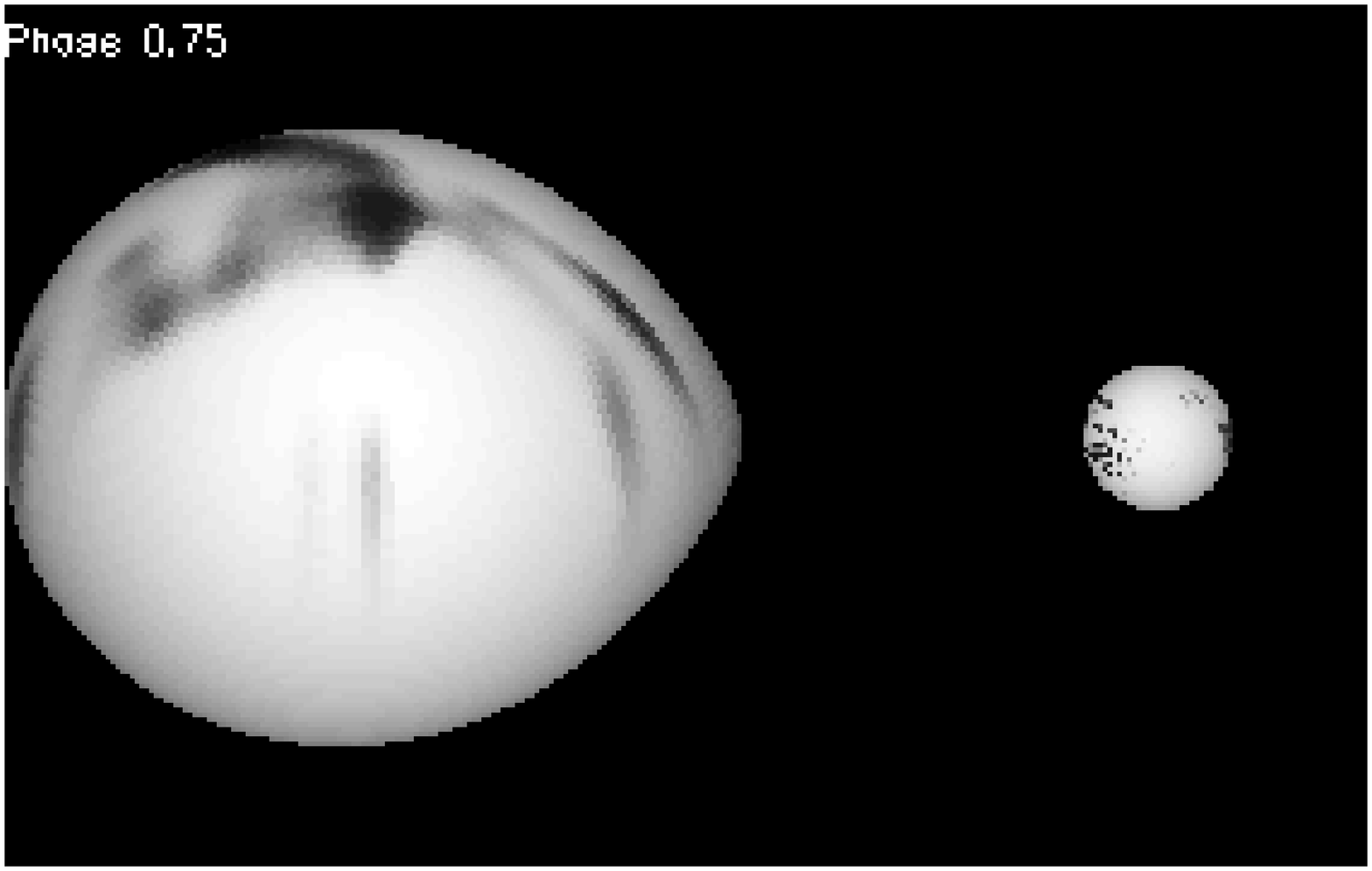}
\caption{Images of SZ Psc at phases 0, 0.25, 0.5 and 0.75, of one orbital cycle (HJD2454067.04--2454071.00).}
\label{fig:images}
\end{figure}

We have presented the first Doppler images of the cooler component of SZ Psc for 2004 November and 2006 September--December. The surface images indicate significant starspot activities for all of the observing seasons. Starspots on the surface of the cooler component were located at various latitudes and longitudes. The presence of pronounced high-latitude or polar spots was revealed by all of our Doppler images. However, the surface maps do not show large polar cap on the cooler component during our observations, which is commonly found on many other active RS CVn-type binaries \citep{str2009}. On the other hand, low-to-mid latitudes of the cooler component of SZ Psc also exhibited numerous starspot groups during the observations. It should be noted that the smearing and vertical elongation of the starspot features near the equator of the star is due to the poor latitude discrimination of Doppler imaging around the equator, which will be worse on stars with high inclination and poor phase sampling \citep{cam1994,bar2004}. The test reconstructions in Appendix A indicate that the lack of a high-latitude spot around phase 0--0.1 in October is most likely due to the phasing, and not to spot evolution. Besides, the high-latitude spot at phase 0.7 is a real feature.

Rapidly rotating cool stars with convective envelop always exhibit starspot activities. The presence of starspots in polar regions of rapidly rotating Sun-like stars, which is not shown on the Sun, is predicted by various numerical simulations \citep{gra2000,mac2004}. Their results suggested that the Coriolis force and the meridional advection are playing important role in locations of starspots on rapid rotators. The K star in SZ Psc has a high rotational velocity and shows pronounced high-latitude and polar starspots, which is consistent with the results of these models. The simulations of the generation and transport of the flux performed by \citet{isik2011} revealed coexistence of high- and low-latitude starspot activities on rapidly rotating main-sequence stars. However, their simulations did not show low-latitude features on K-type subgiants, which have been detected by various observations. They suggested that a possible reason is different meridional flow profiles on these stars, which may produce equatorward transport.

Starspot activities of SZ Psc have been investigated via light-curve modelling by many authors. Their results indicate the spottedness on the surface of the cooler component of SZ Psc in all observing seasons. However, most of these studies only revealed 1--3 large spots on the cooler component for fitting the light curves due to less constraint on spot locations provided by photometric data. Doppler imaging technique can offer us a more detailed distribution of starspots on the cooler component of SZ Psc. As shown in Fig. \ref{fig:mercator}--\ref{fig:images}, our new surface images reveal more complex structure of starspots on the K subgiant component. A number of starspot groups are found to be located at various longitudes during our observations. This is very consistent with the result of \citet{eaton2007}, who found that there should be more than 15 small starspots on the K star to fit line profile distortions and light curves.

From long-term photometric observations, \citet{lan2001} derived spot maps of both components and found several active longitudes on the cooler component. They also revealed two active regions on each component of SZ Psc, facing the other star; the one on the cooler component was long-lived and compact. Our Doppler images also revealed non-uniform longitudinal spot distribution on the cooler component, where starspots were concentrated in several active longitudes, but we do not find any stable active region around the sub-stellar point of the K star during our observations. However, the high-latitude spot around phase 0 seemed to be persistent in most of our images, except for 2006 October, when there is no observation around that phase (see also Fig. A1). The images do not show a clear relationship between the positions of starspots and the stars. This is different from that on another active binary ER Vul, where low- and intermediate-latitude active regions of two components were always located at the hemispheres facing each other probably due to the tidal interactions \citep{xiang2015}. As shown in Fig. \ref{fig:images}, we also find that the hotter component may exhibit starspot activity, which is consistent with the results of \citet{lan2001}. However, due to the low rotational velocity and the non-synchronous rotation of the hotter component, we can not derive a reliable distribution of starspots on this star and do not know how the locations of its starspots are related to the position of the cooler component yet.

The observations in 2006 spanning about three months provide some clues of migrations and evolutions of individual starspots. As shown in Fig. \ref{fig:mercator}, the low-latitude starspot groups around phases 0.1 and 0.35 in September seem to migrate to phases 0 and 0.25 three months later. Besides, the pronounced low-latitude spot group around phase 0.65 in October was faded out in December. However, the different phase coverage also results in different spot patterns, especially for high-latitude features. Hence further Doppler images with complete phase coverage are needed to confirm spot migration and evolution on the cooler component of SZ Psc. \citet{kang2003} suggested that migrations and evolutions of starspots are mainly responsible for the shape changes of the light curves of SZ Psc. Using a two-spot model, they fitted the light curves of SZ Psc obtained in 1978--1981 and found two spots on the cooler component had respectively migrated 65\de\ and 170\de\ in longitude during the observations. However, the two-spot model does not offer any information of individual starspots and thus it is not sure that their results can be attributed to starspot migrations.

Previously, \citet{zhang2008} investigated the chromospheric activity indicators, such as \mbox{He~{\sc i} {\sl D$_{3}$}}, \mbox{Na~{\sc i}} Doublet, H$\alpha$, \mbox{Ca~{\sc ii} IRT}, using the same spectral data sets as ours. The results showed rotational modulation of the chromospheric activity for SZ Psc in 2006. They found that the equivalent widths of several chromospheric lines reached their maximum values at phases 0.25 and 0.75, which indicates two active longitudes on the surface of the cooler component. The emission lines were also strong at phase 0.09. The longitude distribution of starspots revealed by the 2006 images seems to match the rotational modulation of the chromospheric activity. However, since \citet{zhang2008} combined all data spanning about three months, from 2006 September to December, evolutions of starspots might take place during the observations.

We also compare the starspot activities of SZ Psc with that of other similar binary systems. The K1 subgiant component of the well-studied RS CVn-type binary HR 1099 has an orbital period of about 2.8 d. \citet{vog1999} derived Doppler images for HR 1099 from the observing runs between 1981 and 1992. They found the presence of polar cap in the entire observations, which suggests a large, stable polar spot with lifetime longer than 11 years. Our Doppler images of SZ Psc do not show any large polar cap spots which are comparable to that on HR 1099, but show several high-latitude or polar spots on SZ Psc instead. From consecutive Doppler images, \citet{vog1999} and \citet{str2000} revealed a complicated behaviour of starspots on the surface of HR 1099, where several spots at low-to-mid latitudes were migrating in both latitude and longitude and merged with a high-latitude one eventually.

Another similar RS CVn-type system II Peg, which consists of a K2 subgiant and an unseen M dwarf, is one of the most active binaries. The K star exhibits similar latitude distribution of starspots to SZ Psc, although its rotational velocity is relatively slow (~22.6 \kms). The presence of high-latitude and polar starspots as well as low-latitude active regions on the surface of II Peg was revealed by many Doppler images for various observing seasons \citep{gu2003,koc2013,xiang2014}. \citet{ber1998} revealed two long-lived active longitudes, which were separated by half of the rotational period and showed persistent longitudinal migrations. \citet*{ber1999} found the strong correlation between the starspot locations and the strength of the chromospheric emission lines and thus inferred a two-component structure in its chromosphere.

\begin{figure}
\centering
\includegraphics[width=0.47\textwidth]{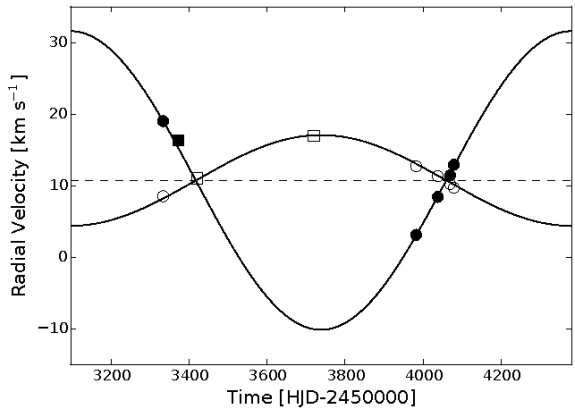}
\caption{Radial velocities of the third body (filled circles) and the binary system (open circles), derived from each data set. The filled square represents the value derived by \citet{gla2008} for the third component and the open squares represent the systemic velocities derived by \citet{eaton2007} for the binary. The lines show the sinusoidal fits.}
\label{fig:rv3}
\end{figure}

Apart from deriving Doppler images, we also detected clear signatures of the third star in the LSD profiles of SZ Psc with high S/N, observed at various orbital phases. According to the strength of the absorption line of the third component during non-eclipse phase, we estimated that the contribution of it is about 5\% of the luminosity of the system, assuming that its line profile is the same as that of G dwarf star. However, the luminosity of the system is reduced by starspots all the time, which leads to overestimate of the contribution of the third component. Our result is in good agreement with that of \citet{eaton2007} , who estimate the contribution of the third component to be 3\%--4\%.

We present the radial velocity curve of the third body and the systemic velocity curve of the binary derived from each data set in Fig. \ref{fig:rv3}. Combining a radial velocity value for the third body derived by \citet{gla2008} in 2005 January and two for the binary system by \citet{eaton2007} in 2005 February and November, we try to find a solution to fit both radial velocity curves simultaneously, assuming a simple, circular orbit. As results, we find a period of $1283 \pm 10$ d and a radial velocity amplitude of $20.9 \pm 0.9$ \kms\ for the third component and $6.3 \pm 0.3$ \kms\ for the binary system (Fig. \ref{fig:rv3}). Therefore we may infer that the mass of the third star is about $0.9 M_{\odot}$. Our results are consistent with those of \citet{eaton2007}, who suggested that the radial velocity amplitude of the binary system is less than 8 \kms\ and the mass of the third component is about 0.9--1$M_{\odot}$. If so, the existence of the third body alone is insufficient to account for the orbital period change of SZ Psc, according to \citet{kal1995}. However, our period estimate is different from that of \citet{eaton2007}, who found a period of 1143 or 1530 d from phasing their data and \citet{pop1988}'s, and excluded other periods. A possible reason is that the outer orbit of SZ Psc is eccentric. Because we assumed a circular orbit, we may have systematic errors in the estimate of the orbital period. It is obvious that our available data were too few to derive a precise orbital solution for the triple system, thus more spectroscopic observations are urgently needed in the future.

\section{Conclusions}

We derived the first Doppler images of the RS CVn-type binary SZ Psc from the spectral data sets obtained in 2004 and 2006 observing seasons. The new Doppler images reveal significant spot activities on the cooler component of SZ Psc, which is consistent with the results of photometric studies. But the distribution of starspots is more complex than that revealed by light-curve modelling. The K subgiant exhibits starspots at various latitudes and longitudes. The images do not indicate any large, stable polar cap, which is different from that of many other RS CVn-type binaries.

Another important result is the detection of clear absorption features contributed by a third component in the high S/N LSD profiles of SZ Psc observed at various orbital phases, which confirms that SZ Psc is a triple system. A preliminary solution of both radial velocity curves of the third star and the binary system indicates that the third component of SZ Psc has a mass of about $0.9 M_{\odot}$ and an orbital period of $1283 \pm 10$ d.

Our results suggest that more high-resolution spectroscopic observations are required to further investigate the starspot activity and the third body of SZ Psc.

\section{Acknowledgements}

This work has made use of the VALD database, operated at Uppsala University, the Institute of Astronomy RAS in Moscow, and the University of Vienna. We would like to thank Profs. Jianyan Wei and Xiaojun Jiang for the allocation of observing time of the Xinglong 2.16m telescope. We are also very grateful to the anonymous referee for helpful comments and suggestions that significantly improved the clarity and quality of this paper. This work is supported by National Natural Science Foundation of China through grants Nos. 10373023, 10773027, 11333006 and U1431114, Chinese Academy of Sciences through project No. KJCX2-YW-T24.

\newpage
\appendix
\section{Reliability test}

Since SZ Psc has a nearly integral-day period, it is difficult to observe this system effectively. In addition, the data sets obtained in 2004 November, 2006 September and 2006 October have large phase gaps owing to the limited observing time. Incomplete phase coverage results in a loss of information of starspots and may cause artefacts in reconstructed images. Therefore we performed tests to show the effect of poor phase sampling on our Doppler images of the cooler component of SZ Psc.

\begin{figure}
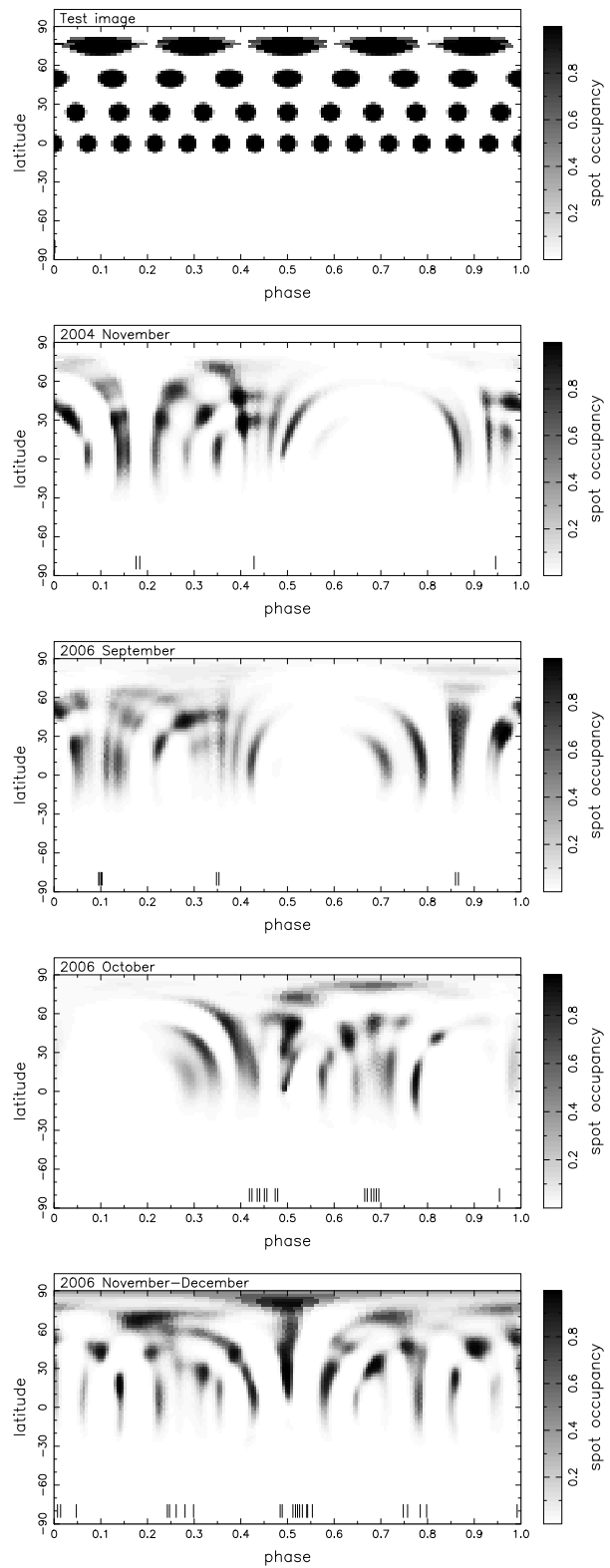

\centering
\includegraphics[angle=270,width=0.46\textwidth]{test.eps}
\includegraphics[angle=270,width=0.46\textwidth]{t0411.eps}
\includegraphics[angle=270,width=0.46\textwidth]{t0609.eps}
\includegraphics[angle=270,width=0.46\textwidth]{t0610.eps}
\includegraphics[angle=270,width=0.46\textwidth]{t061112.eps}
\caption{Effects of incomplete phase coverage. The top panel shows the image that we used for producing artificial data sets with similar S/N to our observations. Others show images derived from the fake data sets.}
\label{fig:phase}
\end{figure}

We produced a fake image covered by many small spots, as shown in the first panel of Fig. \ref{fig:phase}, to generate artificial data sets, which have similar S/N and phase sampling to our real observations. Then we reconstructed Doppler images from these fake data sets and plot the resulting images in Fig. \ref{fig:phase} as well. The results indicate that arc-shaped features in our Doppler images of SZ Psc are spurious, even for 2006 November--December, which has the best phase coverage of our observing seasons. The images of 2004 November, 2006 September and 2006 October show much more pronounced arc-shaped artefacts at large phase gaps. In addition, spot features in these large gaps are absent in the reconstructed images. The distribution of starspots at observed phases is somewhat reliable, whereas the equatorial features are smeared due to the poor latitude discrimination of the Doppler imaging technique at low latitudes and the positions of near-polar features are also inaccurate.

\end{document}